\documentclass[%
reprint,
superscriptaddress,
amssymb,
aps,
floatfix,
]{revtex4-2}

\usepackage{graphicx}
\usepackage{rotating}
\usepackage{amsmath}
\usepackage{bbm}
\usepackage{subfigure}
\usepackage{color}
\usepackage{braket}
\usepackage{tikz}
\usepackage{svg}
\usepackage{hyperref}
\hypersetup{colorlinks, 
	linkcolor={blue!75!black!80!yellow},
	citecolor={blue!75!black!80!yellow}, 
	urlcolor={blue!75!black!80!yellow}
	}

\makeatletter \renewcommand\@make@capt@title[2]{%
\@ifx@empty\float@link{\@firstofone}{\expandafter\href\expandafter{\float@link}}%
\sffamily{\textbf{#1}}\@caption@fignum@sep#2 }

\begin{document}

\title{Simulating polaritonic ground states on noisy quantum devices}

\author{Mohammad Hassan}
\affiliation{Department of Physics, City College of New York, New York, NY 10031, USA}
\affiliation{Department of Physics, The Graduate Center, City University of New York, New York, NY 10016, USA}

\author{Fabijan Pavo\v{s}evi\'{c}}
\email[Electronic address:\;]
{fpavosevic@gmail.com}
\affiliation{Algorithmiq Ltd., Kanavakatu 3C, FI-00160 Helsinki, Finland}
\affiliation{Center for Computational Quantum Physics, Flatiron Institute, 162 5th Ave., New York, NY 10010,  USA}

\author{Derek S. Wang}
\affiliation{Harvard John A. Paulson School of Engineering and Applied Sciences, Harvard University, Cambridge, MA 02138, USA}

\author{Johannes Flick}
  \email[Electronic address:\;]{jflick@ccny.cuny.edu}
\affiliation{Department of Physics, City College of New York, New York, NY 10031, USA}
\affiliation{Department of Physics, The Graduate Center, City University of New York, New York, NY 10016, USA}
\affiliation{Center for Computational Quantum Physics, Flatiron Institute, 162 5th Ave., New York, NY 10010,  USA}

\date{\today}

\begin{abstract}
The recent advent of quantum algorithms for noisy quantum devices offers a new route toward simulating strong light-matter interactions of molecules in optical cavities for polaritonic chemistry. In this work, we introduce a general framework for simulating electron-photon coupled systems on small, noisy quantum devices. This method is based on the variational quantum eigensolver (VQE) with the polaritonic unitary coupled cluster (PUCC) ansatz. To achieve chemical accuracy, we exploit various symmetries in qubit reduction methods, such as electron-photon parity, and use recently developed error mitigation schemes, such as the reference zero-noise extrapolation method. We explore the robustness of the VQE-PUCC approach across a diverse set of regimes for the bond length, cavity frequency, and coupling strength of the H$_2$ molecule in an optical cavity. To quantify the performance, we measure two properties: ground-state energy, fundamentally relevant to chemical reactivity, and photon number, an experimentally accessible general indicator of electron-photon correlation.
\end{abstract}
\date{\today}

\maketitle

\section{Introduction}

Polaritonic chemistry
is a field that exploits strong light-matter interactions for manipulating chemical reactions~\cite{ebbesen2016hybrid,ruggenthaler2018quantum,FlickRiveraNarang2018, garcia2021manipulating,fregoni2022theoretical,li2022molecular,yuen2022polariton,Mandal2023}. These strong light-matter interactions can be created inside micro/nano/picocavities~\cite{frisk2019ultrastrong,benz2016single,baumberg2022picocavities} where the light field strongly couples with a molecular system, giving rise to hybrid light-matter states known as molecular polaritons. By tuning the light field strength inside a cavity, a molecular reaction can then be selectively and non-invasively modified~\cite{thomas2016ground,thomas2019tilting,ahn2023}.

These experimental advances have been accompanied by the development of theoretical methods that can offer insights into the fundamental principles that govern polariton-assisted chemical transformations. Among the various theoretical approaches~\cite{fregoni2018manipulating,climent2019plasmonic,li2020origin,li2020resonance,campos2020polaritonic,fregoni2022theoretical,sanchez2022theoretical,mandal2022theoretical}, a particularly robust class is the quantum electrodynamics (QED) {\it ab initio} methods, such as quantum electrodynamics density functional theory (QED-DFT)~\cite{PhysRevLett.110.233001,ruggenthaler2014quantum,flick2022simple, vu2022enhanced} and quantum electrodynamics coupled cluster methods (QED-CC)~\cite{haugland2020coupled,deprince2021cavity,Pavosevic2021,pavovsevic2022wavefunction,pavosevic2021cavity}, in which both electrons and photons are treated quantum mechanically. Although these methods have provided significant insight into chemical reactions inside a  cavity~\cite{schafer2022shining,deprince2021cavity,vu2022enhanced,pavovsevic2022catalysis,pavosevic2023cavity}, practical implementations of these methods rely on numerous approximations limiting their applicability to weakly correlated molecular systems. Therefore, these methods are expected to fail in the correct description of strongly correlated systems which play a crucial role in transition metal chemistry, photochemistry, catalysis, and bond-breaking processes~\cite{lyakh2012multireference}, among others. Strongly correlated systems can be accurately described with multireference quantum chemistry methods where the corresponding wave function is expanded in terms of excited electronic configurations~\cite{Szalay2011}. However, because the number of these configurations scales factorially with the system size, multireference methods are unsuitable for widespread and black-box applications.

Quantum algorithms offer an alternative way for solving complex chemical problems~\cite{cao2019quantum, bauer2020quantum, head2020quantum}.
One such quantum algorithm for solving the electronic structure problem is the quantum phase estimation (QPE) algorithm~\cite{aspuru2005simulated}. Although the QPE algorithm offers an exponential speedup for the evaluation of eigenvalues of the molecular Hamiltonian over its classical counterparts, it requires very deep circuits comprised of millions of quantum gates~\cite{lanyon2010towards}. Therefore this algorithm exceeds the capacity of currently available noisy quantum devices~\cite{preskill2018quantum}.

An alternative to the QPE algorithm that is more suitable for noisy quantum devices is an iterative hybrid quantum-classical algorithm: the variational quantum eigensolver (VQE)~\cite{peruzzo2014variational}. In the VQE, the quantum device is utilized for the classically intractable parts of computations, such as the quantum state preparation and the ground-state energy measurement, whereas the classical computer is employed for optimization of the parameters that minimize the ground-state energy. By outsourcing the optimization routine to a classical computer, the VQE algorithm exhibits more modest quantum requirements than the QPE algorithm. The VQE algorithm has been successfully implemented on different noisy quantum devices for simulation of molecular systems, including trapped ion devices~\cite{shen2017quantum,hempel2018quantum}, photonic quantum processors~\cite{peruzzo2014variational}, and on superconducting qubits~\cite{o2016scalable,kandala2017hardware,rossmannek2023quantum}.

In practice, the quality of the VQE result depends on the choice of the parametrized wave function ansatz. In its first implementation, the VQE algorithm employed the unitary coupled cluster with singles and doubles (UCCSD) ansatz~\cite{peruzzo2014variational}. The advantages of the UCCSD ansatz are that it is unitary, variational, systematically improvable, and performant for the simulation of strongly correlated systems, in particular for bond-breaking processes~\cite{o2016scalable}. However, because the UCCSD ansatz results in deep quantum circuits along with a large number of wave function parameters, practical implementations of this method on quantum devices require careful optimization of the algorithm, such as term truncation, to minimize quantum resources. 

Encouraged by the success of the VQE algorithm for molecular simulations on noisy quantum devices, as well as by the need for a robust method for simulation of strongly correlated systems confined to a cavity, we have previously developed an ansatz referred to as the quantum electrodynamics unitary coupled cluster with singles and doubles (QED-UCCSD)~\cite{Pavosevic2021}. Therein, we have demonstrated that the QED-UCCSD ansatz for VQE is in excellent agreement with the exact method even in the regions of the potential energy surface where the strong electronic correlations become significant. In the remainder of this work, we will refer to the QED-UCCSD ansatz as the polaritonic unitary coupled cluster (PUCC) ansatz and VQE-PUCC will refer to the method of using the VQE with the PUCC ansatz.

In this work, we study the VQE-PUCC method using realistic noise models and leverage physical symmetries and composite error mitigation strategies to improve the results. The main motivation for this work is to present the formalism, working equations, and technical details for efficient use of quantum devices to simulate systems where electrons and photons are strongly coupled. Among these technical details is the deployment of appropriate error mitigation techniques to precisely resolve ground-state observables, such as the ground-state energies and photon numbers, even in coupling regimes where differences are small. Overall, these results demonstrate the practical feasibility of quantum simulations of mixed fermion-boson systems, such as molecular polaritons, on noisy quantum hardware.

\section{Theory}

The interaction of an embedded molecular system inside an optical cavity and the quantized light field can be described by the Pauli-Fierz Hamiltonian~\cite{ruggenthaler2018quantum}. This Hamiltonian---within the dipole approximation and in the coherent state basis~\cite{haugland2020coupled} for a single-mode cavity in second quantization notation---reads as
\begin{equation}
\begin{aligned}
    \label{eqn:PF_Hamiltonian_CSB}
    \hat{H}&=h^p_q a_p^q + \frac{1}{4}\bar{g}^{pq}_{rs} a_{pq}^{rs}\\\newline& + \omega b^{\dagger}b-\sqrt{\frac{\omega}{2}}(\boldsymbol{\lambda} \cdot \Delta\boldsymbol{d})(b^{\dagger}+b)+\frac{1}{2}(\boldsymbol{\lambda} \cdot \Delta\boldsymbol{d})^2.
\end{aligned}
\end{equation}
The first two terms in this equation correspond to the electronic Born-Oppenheimer Hamiltonian (note that extensions to include non-adiabatic effects are also possible~\cite{Hammes-Schiffer19_338,Hammes-Schiffer20_4222,pavosevic2021multicomponent}), where $a_{p_1p_2...p_n}^{q_1q_2...q_n}=a_{q_1}^{\dagger}a_{q_2}^{\dagger}...a_{q_n}^{\dagger}a_{p_n}...a_{p_2}a_{p_1}$ are the second-quantized electronic excitation operators expressed in terms of fermionic creation ($a^{\dagger}$) and annihilation ($a$) operators. Additionally, $h^p_q=\langle q |\hat{h}|p \rangle$ is a matrix element of the one-electron core
Hamiltonian and $\overline{g}^{pq}_{rs}=g^{pq}_{rs}-g^{qp}_{rs}=\langle rs|pq \rangle - \langle rs|qp \rangle$ is the antisymmetrized two-electron Coulomb
repulsion tensor element. Furthermore, $p, q, r, s,...$ indices denote general electronic
spin orbitals, $i,j,k,l,...$ indices denote occupied electronic spin orbitals, and $a,b,c,d,...$ indices denote
unoccupied (virtual) electronic spin orbitals. The third term of Eq.~\eqref{eqn:PF_Hamiltonian_CSB} represents the Hamiltonian of the cavity photon mode with the frequency $\omega$. There, the operators $b^{\dagger}$ and $b$ are bosonic creation and annihilation operators, respectively. The fourth term accounts for the interactions of electronic and photonic degrees of freedom. Finally, the last term corresponds to the dipole self-energy.  
Within the last two terms, $\boldsymbol{\lambda}=(\lambda_x, \lambda_y, \lambda_z)$ with $\lambda=|\boldsymbol{\lambda}|$ is the coupling strength vector and $\Delta\boldsymbol{d}=\boldsymbol{d}-\langle\boldsymbol{d}\rangle$ is the total molecular dipole moment operator subtracted by its expectation value that arises due to the coherent-state basis transformation~\cite{haugland2020coupled}. Note that the total molecular dipole moment is defined as $\boldsymbol{d}=\boldsymbol{d}_{\text{electronic}}+\boldsymbol{d}_{\text{nuclear}}$.

\begin{figure*}[htp!]
\centering
\centerline{\includegraphics[width=0.9\textwidth]{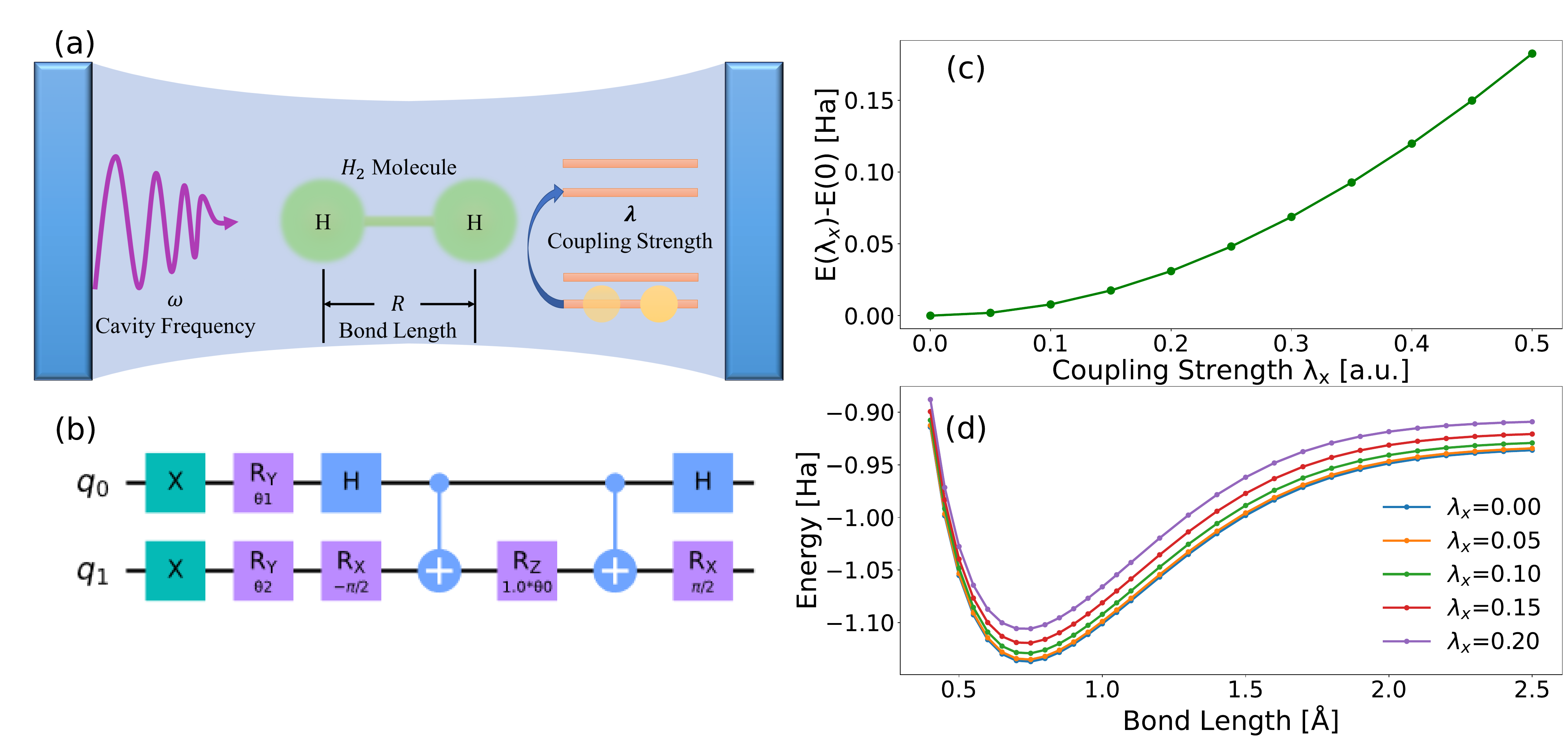}}
\caption{\textbf{(a)} Schematic of an H$_2$ Molecule in an optical cavity. A cavity mode of frequency $\omega$ interacts with the electronic system of H$_2$ with bond length $R$ via the electron-photon coupling strength $\lambda$. \textbf{(b)} The reduced ansatz circuit obtained by using a Bravyi-Kitaev mapping and tapering off a qubit using Z$_2$ symmetries. \textbf{(c)} and \textbf{(d)} show FCI calculations for H$_2$ interacting with a single photon mode. \textbf{(c)} The energy difference of the system between the coupling strength $\lambda_x$ and $\lambda=0$~a.u. ($\omega=2$~eV, $R_0=0.735 \textnormal{\AA}$). \textbf{(d)} Dissociation curves at increasing values of $\lambda_x$ ($\omega=2$~eV).}
\label{fig:motivation_plot}
\end{figure*}

Within the VQE algorithm, the ground state energy of a system inside a cavity ($E_{\text{QED}}$) is determined by variationally minimizing the energy functional
\begin{equation}
    \label{eqn:QED-VQE-Funtional}
     E_{\text{QED}}=\underset{\theta_{\mu,n}}{\text{min}} \langle\Psi_{\text{QED}}(\theta_{\mu,n})|\hat{H}|\Psi_{\text{QED}}(\theta_{\mu,n})\rangle
\end{equation}
with respect to wave function parameters $\theta_{\mu,n}$. In this equation, $|\Psi_{\text{QED}}(\theta_{\mu,n})\rangle$ is the trial wave function that depends on parameters $\theta_{\mu,n}$. In the case of the PUCC method, the wave function is defined as 
\begin{equation}
    \label{eqn:PUCC}
    |\Psi_{\text{PUCC}}\rangle=e^{\hat{T}-\hat{T}^\dagger}|0^{\text{e}}0^{\text{ph}}\rangle,
\end{equation}
where $|0^{\text{e}}0^{\text{ph}}\rangle=|0^{\text{e}}\rangle\otimes|0^{\text{ph}}\rangle$ corresponds to the reference quantum electrodynamics Hartree-Fock (QED-HF) wave function~\cite{haugland2020coupled} defined in terms of an electronic Slater determinant ($|0^{\text{e}}\rangle$) and a zero-photon state ($|0^{\text{ph}}\rangle$). Because the interaction of the quantum particles (i.e. electrons and photons) is accounted for through a mean-field potential, correlation effects are not included in the QED-HF method. In Eq.~\eqref{eqn:PUCC}, $\hat{T}$ is the cluster operator
\begin{equation}
    \label{eqn:QED-T}
    \hat{T}=\sum_{\mu,n}\theta_{\mu,n}a^{\mu}(b^{\dagger})^n,
\end{equation} 
where $a^{\mu}=a_{\mu}^{\dagger}=\{a_{i}^{a},a_{ij}^{ab},...\}$ is a set of single, double, and higher electronic excitation operators, $\mu$ is an excitation manifold, and $n$ is the number of photons. Therefore, action of the operator $a^{\mu}(b^{\dagger})^n$ on the QED-HF reference will produce the excited configuration
\begin{equation}
    \label{eqn:T_on_QED-HF}
    a^{\mu}(b^{\dagger})^n|0^{\text{e}}0^{\text{ph}}\rangle=\sqrt{n!}|\mu\rangle\otimes|n\rangle
\end{equation}
that accounts for the correlation effects between quantum particles. Because of the unitary form of the wave function~\cite{Pavosevic2021}, the PUCC method is suitable for quantum computations, since quantum computers implement unitary operators via quantum gates~\cite{peruzzo2014variational,romero2018strategies,cao2019quantum,bauer2020quantum,head2020quantum}. Truncation of the excitation cluster operator up to a certain electronic excitation manifold and the number of photons establishes the systematically improvable PUCC hierarchy. In cases where the electronic excitation rank equals the number of electrons, the method is said to be exact. In this work, we consider the truncated cluster operator defined as
\begin{equation}
    \label{eqn:QED-T_trun}
    \hat{T}=\theta^{i,0}_a a_i^a+\theta^{0,1}b^\dagger+\frac{1}{4}\theta^{ij,0}_{ab} a_{ij}^{ab}+\theta^{i,1}_aa_i^ab^\dagger+\frac{1}{4}\theta^{ij,1}_{ab} a_{ij}^{ab}b^\dagger
\end{equation}
where up to one photon and up to two electrons interact with each other~\cite{haugland2020coupled}. In the original proposal, this method was coined as QED-UCCSD-1~\cite{Pavosevic2021}. Importantly, the QED-UCCSD-1 method is exact for systems comprising of two electrons interacting with at most one photon. One such system that we study in this work is the interaction of a single photon with the two electrons of an H$_2$ molecule in an optical cavity, as illustrated in Fig.~\ref{fig:motivation_plot} (a).

An implementation of the VQE-PUCC method on a quantum device follows the standard procedure~\cite{peruzzo2014variational,tilly2022variational}. In the first step, the second-quantized Hamiltonian from Eq.~\eqref{eqn:PF_Hamiltonian_CSB} is constructed using the QED-HF spin orbitals. In the second step, the fermionic and bosonic operators that occur in both the Hamiltonian and cluster operators are converted into the $I,X,Y,Z$ Pauli basis. Due to different spin statistics between electrons (fermions) and photons (bosons), their corresponding creation/annihilation operators are mapped to the qubit basis differently. The fermionic creation and annihilation operators are converted into the qubit basis with Jordan-Wigner mapping~\cite{jordan1993algebraic} or with a more economic Bravyi-Kitaev mapping~\cite{bravyi2002fermionic,seeley2012bravyi}. The latter approach allows the removal of two qubits, thereby significantly reducing the quantum hardware requirements~\cite{bravyi2017tapering}. The bosonic creation and annihilation operators can be converted to the qubit basis by the direct boson mapping~\cite{somma2003quantum,veis2016quantum,Macridin2018} in which the state with $n$ bosons is mapped to $n+1$ qubits. With the direct boson mapping, the bosonic creation operator is transformed as
\begin{equation}
    \label{eqn:Boson-mapping_creation}
        \begin{aligned}
            b^{\dagger}=\sum_{j=0}^{n-1}\sqrt{j+1}\sigma_{+}^{j}\sigma_{-}^{j+1},
        \end{aligned}
\end{equation}
where $\sigma_{\pm}=1/2(X\pm iY)$. Moreover, we will also calculate the photon number operator that is transformed as 
\begin{equation}
    \label{eqn:Number-operator-mapping}
        \begin{aligned}
            b^{\dagger} b=\sum_{j=0}^{n}j \frac{Z^{j}+I}{2}.
        \end{aligned}
\end{equation}
Following the conversion of second-quantized operators to the qubit basis, state preparation with the PUCC ansatz, and energy measurement on a noisy quantum computer, the optimization of the wave function parameters $\theta_{\mu,n}$ is performed on a classical computer. This step is repeated until the convergence in energy is achieved.

For electron-photon systems, we can exploit an additional symmetry to reduce the number of necessary qubits in the simulation. After applying the Bravyi-Kitaev mapping that results in two fewer qubits compared to the Jordan-Wigner mapping, we can taper off an additional qubit by exploiting the total parity operator, which is a $\mathcal{Z}_2$ symmetry~\cite{bravyi2017tapering,whitfield2020tapering}. For the electron-photon system studied here, the total parity operator $\mathcal{P} = \mathcal{P}_\text{e}\cdot \mathcal{P}_\text{ph}$
is the product of the electronic parity $\mathcal{P}_\text{e}$ and the photon parity operator $\mathcal{P}_\text{ph}$ that is defined as  
\begin{align}
\mathcal{P}_\text{ph} &= \exp\left(-i \pi b^\dagger b \right).    
\end{align}
Therefore, using the Bravyi-Kitaev mapping, direct boson mapping, and the parity symmetry, we can describe the system by $n-3$ qubits compared to $n$ qubits needed in the Jordan-Wigner and direct boson mapping alone. Figure~\ref{fig:motivation_plot} (b) shows the ansatz after undergoing a Bravyi-Kitaev mapping and qubit tapering.

\begin{table}[htp!]

\centering
\caption{Dimensions of the cluster excitation operator for $\text H_2$ in a cavity for different fermion-to-qubit mappings. By leveraging Z$_2$ symmetries, we reduce the circuit to two qubits and two entangling CNOT gates.}
\begin{tabular}{|l||l|l|}
\hline
Mapping & Qubits & CNOTs\\
\hline
\hline
Jordan-Wigner (JW) & $5$ & $96$ \\
\hline
Brayvi-Kitaev (BK) & 3 & 10 \\
\hline
BK + Tapering & 2 & 2 \\
\hline
\end{tabular}
\end{table}

\section{Error Mitigation}

Experimentally relevant electron-photon coupling strengths result in comparatively small differences in observables. For instance, ground-state energy shifts between experimentally realizable ranges of $\lambda$ are expected to be on the order of meV~\cite{pavosevic2021cavity}. As an example, Fig.~\ref{fig:motivation_plot} (c) shows full configuration interaction (FCI) calculations for the energy difference between the ground state energy of a H$_2$ molecule outside the optical cavity (i.e. $\lambda = 0$~a.u.) and the ground state energy for finite values of $\lambda_x$ (where $\omega=2$~eV).
These results give an estimate of how high the resolution of the noisy VQE approach must be to resolve energy differences between $\lambda_x$. For example, the energy difference between $\lambda_x=0$~a.u. and $\lambda_x=0.1$~a.u. in Fig.~\ref{fig:motivation_plot} (c) is 7.8 mHa. Similarly, the FCI calculations in Fig.~\ref{fig:motivation_plot} (d) shows how the bond-length dependent energy, or dissociation curve, also shifts on the order of mHa across the relevant range of $\lambda_x$. We find the equilibrium bond length shifting from 0.735~\textnormal{\AA} for $\lambda_x=0$~a.u. to 0.726~\textnormal{\AA} for $\lambda_x=0.2$~a.u. Additionally, we find for the energy difference between $\lambda_x=0$~a.u. and $\lambda_x=0.05$~a.u. a value of  1.96 mHa, which is close to the standard of chemical accuracy of 1.6~mHa ($\sim$ 43~meV) in electronic-structure calculations. Thus, these small energy differences in the weak coupling regime, and in particular if $\lambda < 0.05$~a.u., set an additional criterion for the precision of the quantum simulation. Not only must the results be within the criteria for chemical accuracy, but the noise level must fall within the energy difference to correctly resolve it.

To resolve these differences on small, noisy quantum devices in practice, we employ error mitigation. In this work, we compose several methods: readout-error mitigation~\cite{Nation2021readout}, zero-noise extrapolation (ZNE)~\cite{Temme2017zne, Li2017zne, GiurgicaTiron2020zne}, and reference-state zero-noise extrapolation (rZNE)~\cite{Tzu-Chieh2023hardware}, which is a combination of reference-state (RS) error mitigation~\cite{Tzu-Chieh2023hardware} and ZNE~\cite{Temme2017zne, Li2017zne, GiurgicaTiron2020zne}.
 
The key idea behind ZNE is that the application of a unitary gate, succeeded by the application of its inverse, produces an identity for a noiseless device. Thus, if the whole circuit $U$ is unitary, its copy $U$ and its inverse $U^{-1}$ can be appended to the end of the circuit, i.e. $U=UU^{-1}U$. On a noisy device, this procedure has the effect of amplifying the noise of the original circuit. 
The degree of noise amplification is defined by the noise factor $m$, where $m=2n+1$ and $n$ is the number of times that a copy of the circuit and its inverse is applied, i.e. $U=U(U^{-1}U)^n$. The expectation value of the unitary is measured at different noise factors, and then these noisy expectation values are used to extrapolate the gate-error-free expectation value at a noise factor of 0.

To further reduce errors, we rely on RS error mitigation. Here, the noise of the quantum device is treated as an operation which rescales the exact expectation value into the measured expectation value, i.e., 
\begin{align}
\bra{\boldsymbol{\vec{\theta}}}\hat{O}\ket{\boldsymbol{\vec{\theta}}}_{\mathrm{noisy}}=r\bra{\boldsymbol{\vec{\theta}}}\hat{O}\ket{\boldsymbol{\vec{\theta}}}_{\mathrm{exact}}
\label{eq:rescale}
\end{align}
where $\hat{O}$ is an arbitrary operator, $\boldsymbol{\vec{\theta}}$ are the parameters of the ansatz, and $r$ is the rescale factor. To obtain these rescaling factors, this approach makes use of reference states: states for which the exact expectation values are known or can be constructed noise-free. For the electron-photon systems discussed here, we can use the QED-HF state, where all parameters in the PUCC ansatz are set to 0, as a reference state. The rescale factor is then the ratio of the measured QED-HF energy on the quantum device to the numerically exact QED-HF state obtained from standard electronic-structure packages. The rescale factors can then be used in Eq.~\ref{eq:rescale} to obtain the unbiased expectation value in any arbitrary state. The underlying assumption here is that the noise of the quantum device is only weakly dependent on the specific ansatz parameters. RS mitigation can be combined with other error mitigation techniques, so long as the same combination of mitigation techniques is used for measuring in both the reference state and the desired state.

The rZNE mitigation strategy works by using the ZNE fitting functions of both the reference state and the desired state to calibrate the noisy expectation value. This method is based on the assumption that the noise between two qubits can be modeled as a depolarizing channel, an assumption which holds well for IBM's devices under Pauli twirling~\cite{shao2023depolarizing}. The depolarizing channel can be mathematically expressed as such: $\rho_2 \rightarrow (1-p_m)\rho_2 + p_m\frac{I\bigotimes I}{4}$, where $\rho_2$ is the density matrix representing the state of the two qubits, $p_m$ is the depolarizing probability given by $p_m = 1-e^{-gm}$, $m$ is the noise factor, and $g$ is the device and gate-dependent exponential decay constant that governs how drastically the noise is amplified at a given noise factor. This function is chosen such that if the noise factor becomes large, the density matrix becomes maximally mixed, while when the noise factor goes to zero, the density matrix remains unchanged. It follows that when applying ZNE, the fitting function has the following exponential form, i.e. $f(m)=Ee^{-gm}$, where $E$ is the energy being measured.
To combine the RS mitigation with ZNE, we apply ZNE both when measuring the reference state and when running the VQE algorithm. We obtain the ZNE fitting function for both as such: $f_r(m)=a_r e^{-g_r m}+c_r$ for the reference state and $f_{e}(m)=a_{e}e^{-g_{e}m}+c_{e}$ for the VQE result, where the subscripts $r$ and $e$ denote fitting parameters for the reference state and VQE result, respectively. We note that at zero-noise ($m=0$), $f_r(0)=E_{\mathrm{QED-HF}}=a_r+c_r$ and $f_{e}(0)=E_{\mathrm{unbiased}}=a_e+c_e$, where $E_{\mathrm{QED-HF}}$ is the QED-HF energy and $E_{\mathrm{unbiased}}$ is the energy obtained from the VQE with zero gate noise. We note that in practice, when applying ZNE, the fitting functions contain the additive parameters $c_r$ and $c_e$ for the reference and VQE energies, respectively. Since the energies at $m=0$ still contain noise, we can rescale as such: $E_{\mathrm{QED-HF}}=a_r\cdot r+c_r$ and $E_{\mathrm{unbiased}}=a_e\cdot r+c_e$, where $r$ is the rescale factor. We combine the expressions for the rescale factors with the fitting functions to obtain the following energy expression:
\begin{align}
E_{\mathrm{rZNE}}=a_{e}\frac{E_{\mathrm{QED-HF}}-c_r}{a_r}+c_{e}.
\label{eq:rZNE}
\end{align}
In Eq.~\ref{eq:rZNE}, $E_{\mathrm{rZNE}}$ is the rZNE mitigated VQE energy. We combine rZNE with readout error mitigation by mitigating the readout error for the energy at each noise factor for both the reference state and VQE measurements.

To further reduce the error, we found that it is possible to map the occupied states to $\ket{0}$, which is less susceptible to processes leading to in particular amplitude damping~\cite{nielsen2010quantum}.
For the H$_2$ molecule, there are two electrons; therefore, two of the spin orbitals must be occupied. In the ansatz circuit, this is done by initializing the qubits in the $\ket{1}$ state by operating each qubit with an X-gate, as shown in Fig~\ref{fig:motivation_plot} (b). Since the evolution of the $\ket{1}$ state to the $\ket{0}$ state causes errors in the final result, it is generally ideal to start off in the $\ket{0}$. We avoid initializing the qubits in the $\ket{1}$ state by multiplying the coefficients of each term in the mapped Hamiltonian by $(-1)^k$, where $k$ is the number of Z-gates that appear in that term. Table \ref{tab:Xgate_table} compares the results between an ansatz with X-gates to the ansatz without X-gates. From this, we find that the ansatz with X-gates gives a result with a percent error of 3.2709\%, while the ansatz without X-gates gives a result with a percent error of 2.2997\%. This result shows that indeed removing the X-gates gives a small advantage in accuracy while also reducing computational resources.

\begin{table}[h]
\centering
\caption{\label{tab:Xgate_table} VQE results for polaritonic H$_2$ with $\omega=2$~eV and $\lambda_x=0.1$ a.u. at equilibrium bond length. We compare the results for when the ansatz is initialized in the $\ket{1}$ state (by applying X-gates to the beginning of the ansatz) and $\ket{0}$. For each case, the result is the average of twenty VQE runs. The FCI energy in this case is -1.1295 Ha. The last column shows the standard deviation.}
\end{table}
\begin{tabular}{|l|l|l|l|}
\hline
\textbf{Ansatz} & \textbf{Energy [Ha]} & \textbf{\% Error} & \textbf{Std Dev [Ha]} \\ \hline
$\ket{1}$ & -1.0925 & 3.2709 & 0.0019\\ \hline
$\ket{0}$ & -1.1035 & 2.2997 & 0.0016\\ \hline
\end{tabular}

\section{Technical Details}

In this work, we run all simulations using the Qiskit platform~\cite{Qiskit}. Qiskit offers tools called ``Fake Devices" for simulating the noise models of their devices. In this work, we use the noise model of the IBM Cairo device. All of the simulations are run with twenty thousand shots. For each data point, ten VQE simulations are run, and the average result of these ten runs is plotted. The error bars shown are the root-mean-square error (RMSE) of these ten runs.

Readout-error mitigation is applied using the M3 package~\cite{Nation2021readout}. We use M3 by sampling the calibration data from two physical qubits on the fake device (which can also be done on the real device) and then mapping the virtual qubits of our system to those same physical qubits. The measurement of the system is then corrected by the calibration data during every function evaluation of the VQE loop.

In implementing ZNE, we only scale the CNOT gates. We use scale factors of 1, 3, 5, 51, 101, and 201. Using these scale factors allows us to properly fit the measured expectation values to an exponential so that we can obtain the fitting parameters for use in Eq.~\ref{eq:rZNE}. ZNE is applied during each objective function evaluation. Unlike readout-error mitigation and ZNE, RS mitigation, and rZNE are only applied to the final converged VQE results. When calculating the rescale factors for reference state mitigation and the reference exponential fit parameters for rZNE, we measure the noisy QED-HF state fifty times and take the average.

\section{Results} \label{results}

\begin{figure}[htp!]
\centering
\centerline{\includegraphics[width=0.47\textwidth]{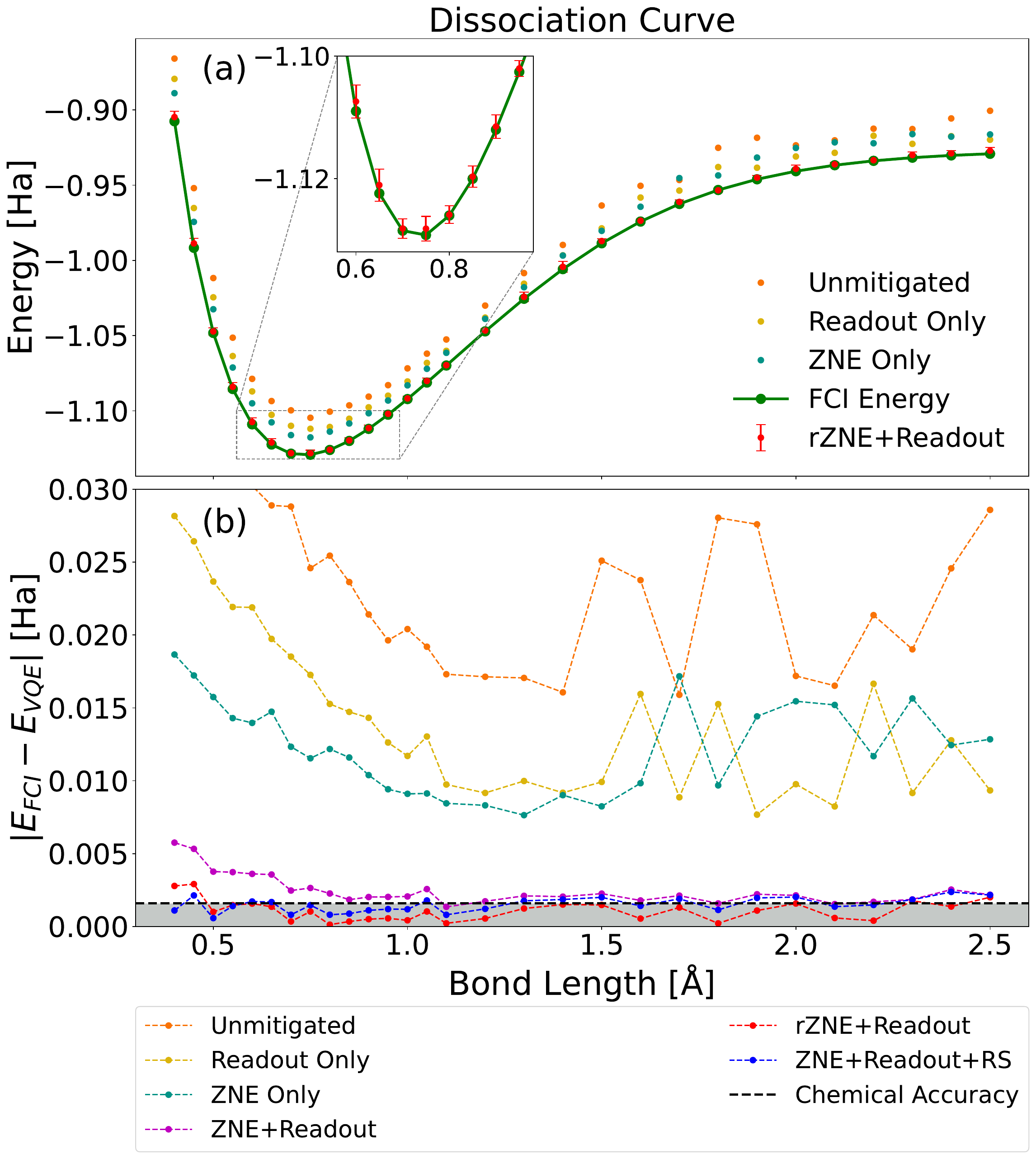}}
\caption{\textbf{(a)} Potential energy surface for H$_2$ coupled to a single photon mode. Results are shown for different levels of error mitigation. The inset shows the error bars near equilibrium. \textbf{(b)} The error relative to the FCI energies for different levels of error mitigation. The gray shaded region is the region in which results are chemically accurate. The first few points for the unmitigated energies are cutoff due to high errors. The results obtained using combinations of ZNE, readout-error mitigation, and RS mitigation are not shown in \textbf{(a)} due to their close overlap with the rZNE results, but their associated errors are shown in \textbf{(b)}. Chemical accuracy can be achieved with combinations of ZNE, readout-error mitigation, and RS mitigation, and can consistently be achieved with rZNE.}
\label{fig:diss}
\end{figure}

We now discuss the results obtained using the IBM Cairo simulator. Figure~\ref{fig:diss} (a) shows the dissociation curve of the H$_2$ molecule inside an optical cavity. In the plot, we show the ground-state energy changing with different bond lengths, where the bond length is defined by the distance between the two hydrogen atoms. In this case, the cavity frequency is chosen as $\omega=2$~eV and the coupling strength is chosen as $\lambda_x= 0.1$~a.u., with the coupling strength in the other directions set to zero. These parameter values are chosen in agreement with previous work~\cite{Flick2018,Pavosevic2021}. The VQE result is plotted for different levels of error mitigation.

Figure~\ref{fig:diss} (b) shows the relative error as the absolute value of the difference between the FCI energy and the VQE energies in Fig.~\ref{fig:diss} (a). We find that we are able to achieve chemical accuracy for many of the bond lengths when combining rZNE with readout-error mitigation and when combining ZNE with readout-error mitigation and RS mitigation. However, for the combination of readout-error mitigation with ZNE, most of the results are just outside of chemical accuracy. This shows quantum computing is a promising avenue for studying electron-photon problems, even in the near term, due to the unique advantage that electron-photon reference states have for enhancing error mitigation.

\begin{figure}[!htp]
\centering
\centerline{\includegraphics[width=0.47\textwidth]{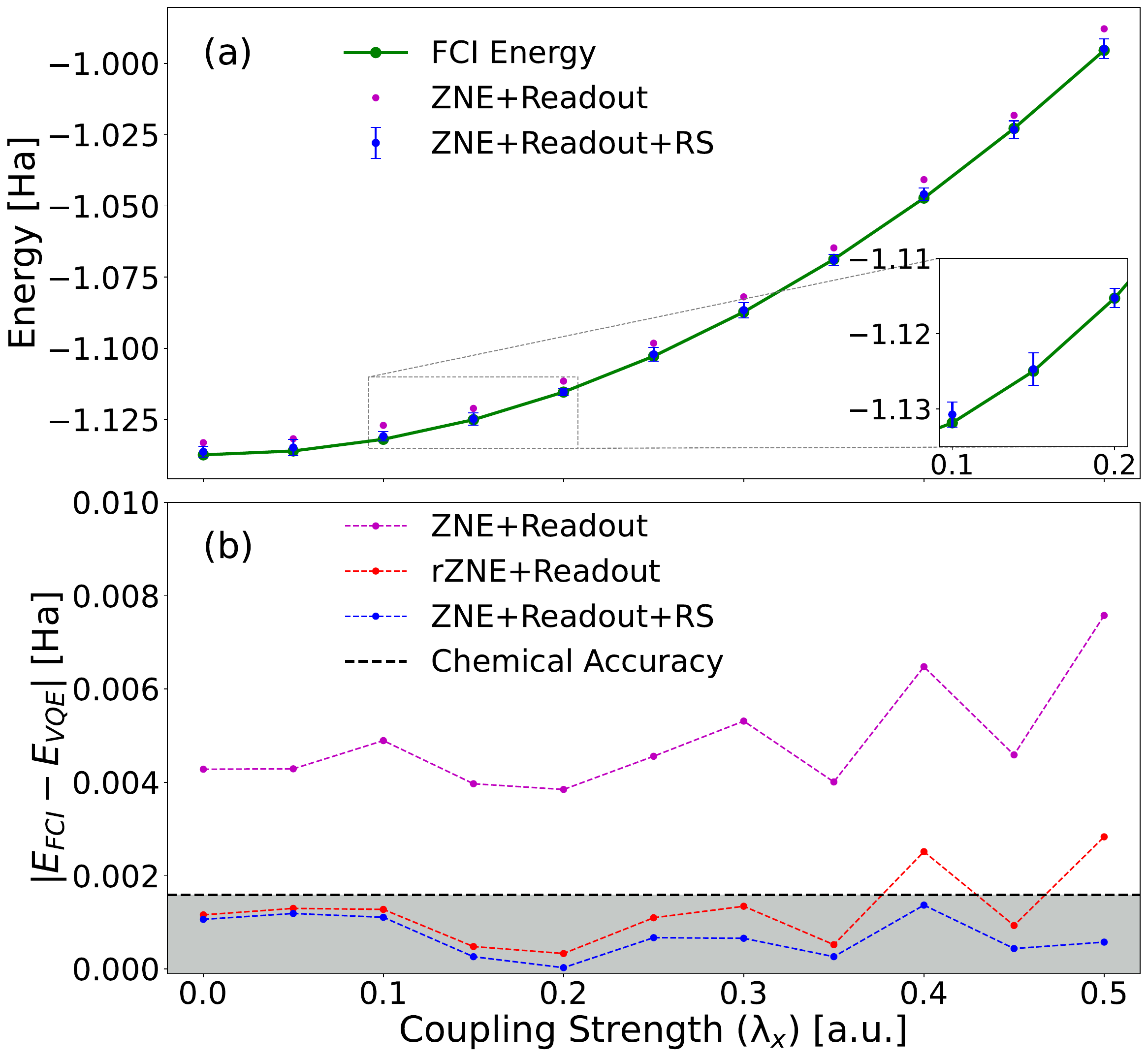}}
\caption{\textbf{(a)} Energy with respect to coupling strength in the x-direction for H$_2$ coupled to a single photon in a single photon mode. Results are shown for different levels of error mitigation. \textbf{(b)} The error relative to the FCI energies for different levels of error mitigation. The shaded region is the region in which results are chemically accurate.}
\label{fig:EnergyLambda}
\end{figure}

In the following, we further explore the robustness of the VQE-PUCC approach across various coupling strengths. To quantify the performance, we measure two properties: the ground-state energy, which is fundamentally relevant to chemical reactivity, and the photon number, which is a general indicator of how correlated the electron-photon system is and can be monitored experimentally.

Thus, Fig.~\ref{fig:EnergyLambda} (a) shows the VQE results for the energy at different coupling strengths for photon polarization along the x-direction. The cavity frequency is set to $\omega=20$~eV. The bond length at each data point is the equilibrium bond length for the given $\omega$ and associated $\lambda_x$. The purpose of the inset is to show the scale of the error bars.

Figure~\ref{fig:EnergyLambda} (b) shows the relative error as the absolute value of the difference between the FCI energy and the VQE energies in Fig, \ref{fig:EnergyLambda}a. Unlike Fig.~\ref{fig:diss}, where rZNE generally performs better than RS mitigation, in this case, RS mitigation performs better than rZNE. In fact, RS mitigation gives chemically accurate results for points.

\begin{figure}[!htp]
\centering
\centerline{\includegraphics[width=0.47\textwidth]{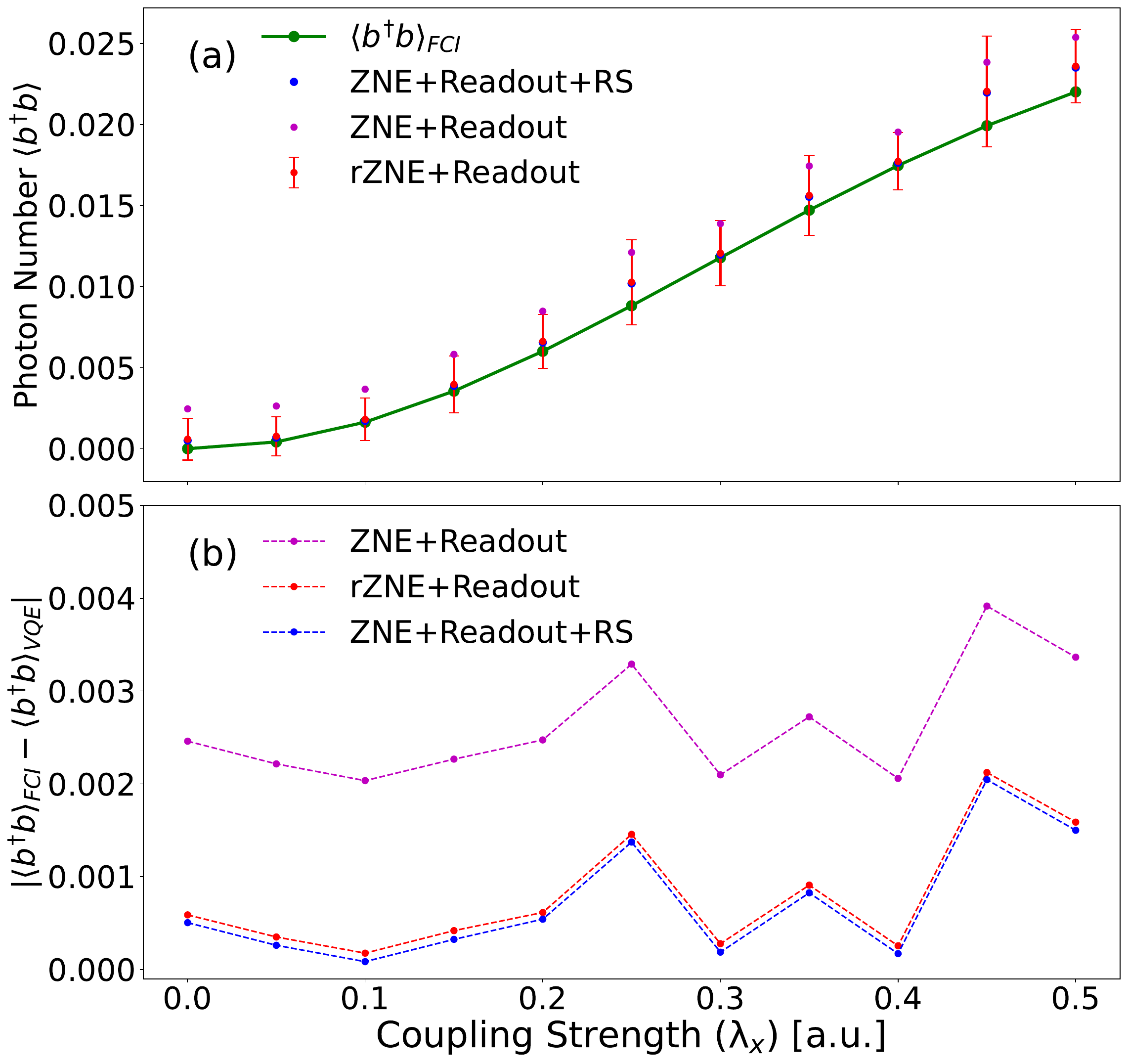}}
\caption{\textbf{(a)} Average photon number with respect to coupling strength in the x-direction for H$_2$ coupled to a single photon in a single photon mode. Results are shown for different levels of error mitigation. \textbf{(b)} The error relative to the FCI average photon number for different levels of error mitigation.} 
\label{fig:photonNum}
\end{figure}

Figure~\ref{fig:photonNum} (a) shows the average photon number. The average photon number is calculated as $\braket{b^{\dagger}b} = \frac{1}{2}(1- \braket{ZZ} )$, where $\braket{b^{\dagger}b}$ is the average photon number and $ZZ$ is the operator that represents the application of a Pauli-Z gate on each of the two matrices. The expectation value is taken at the ground state of the Hamiltonian. To obtain the average photon number in the ground state on a quantum computer as shown in Fig.~\ref{fig:photonNum} (a), we use the $ZZ$ expectation value at the optimal points found by the VQE. These are the same optimal points that give the energies in Fig.~\ref{fig:EnergyLambda} (a). The reference exponential fit parameters for rZNE are found by measuring $\braket{b^{\dagger}b}$ in the QED-HF state at different scale factors. Since the exact result for $\braket{b^{\dagger}b}$ in the QED-HF state is zero, then calculating the rescale factor would result in a division by zero. What we do instead is measure $\braket{ZZ}$ in the QED-HF state. Since the exact expectation value of the $ZZ$ operator is one, the rescale factor is the noisy measurement of $\braket{ZZ}$ in the QED-HF state. We use this rescale factor to rescale the noisy measurement of $\braket{ZZ}$ in the optimal state. We find that even though the optimal points found by the VQE allow for convergence of the energy with small error bars, this does not necessarily mean that the photon number at those optimal points will have similarly small error bars. Simulating the system at a cavity frequency of 20~eV allows us to simulate photon numbers which are much higher than the noise. Though such a high cavity frequency might be hard to realize in a lab, it allows us to demonstrate the ability of the VQE to simulate observables other than energy. We observe that even though the combination of readout-error mitigation and RS mitigation is more accurate than readout-error mitigation and rZNE when calculating the energy (Fig.~\ref{fig:EnergyLambda}), rZNE is more accurate than RS mitigation when calculating the average photon number.

\begin{figure}[htp!]
\centering
\centerline{\includegraphics[width=0.47\textwidth]{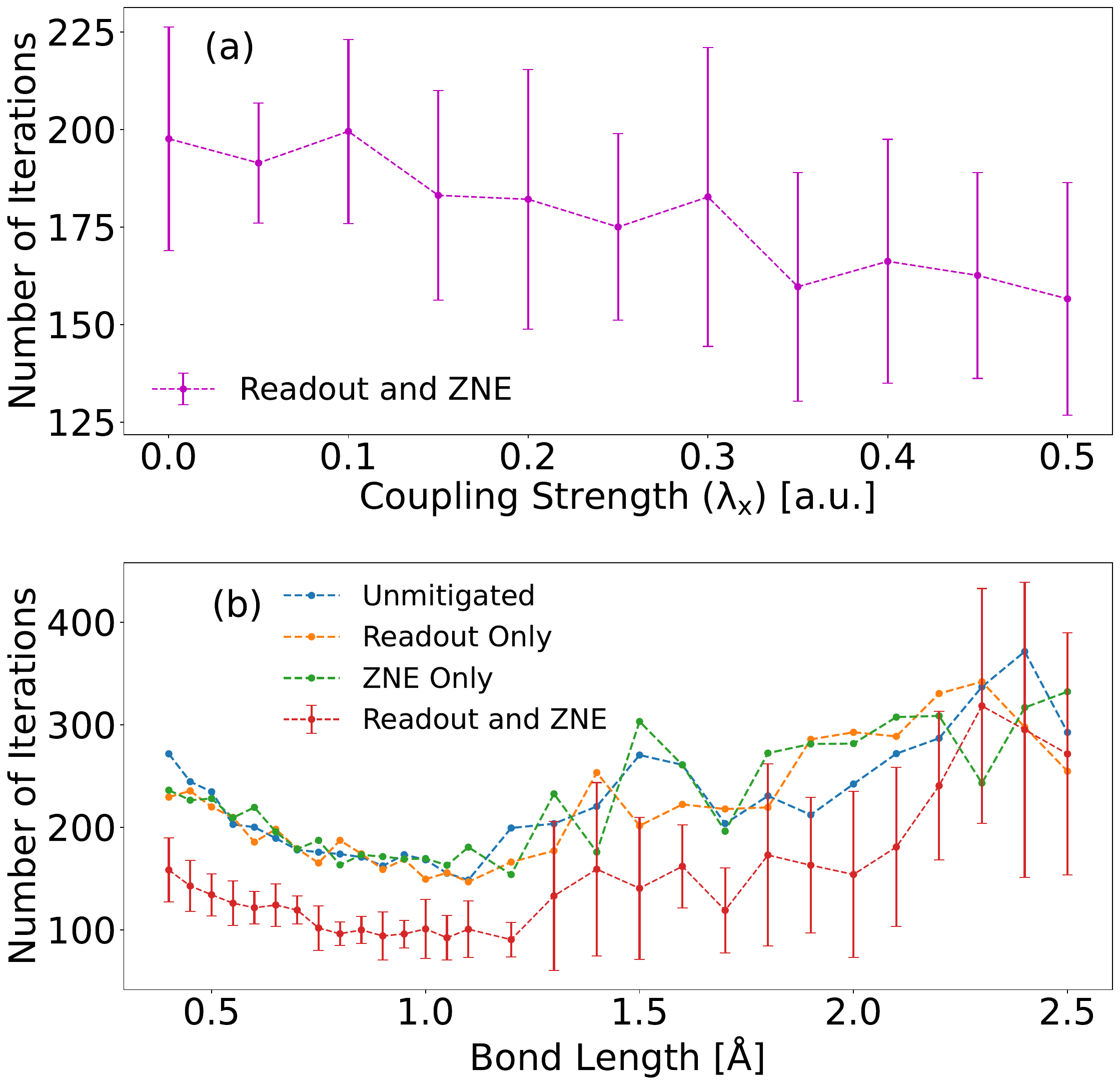}}
\caption{\textbf{(a)} The average number (n=10) of iterations that it takes for the VQE to converge vs. the coupling strength in the x-direction ($\lambda_x$). \textbf{(b)} The average number (n=10) of iterations that it takes for the VQE to converge vs. the bond length. In both subplots, the error bars are the standard deviation of the iterations.}
\label{fig:convergence_plots}
\end{figure}

The convergence data for the VQE simulations are shown in Fig. \ref{fig:convergence_plots}. Fig.~\ref{fig:convergence_plots} (a) shows the average number of iterations for the results shown in Fig.~\ref{fig:EnergyLambda} (a). For each $\lambda_x$, we average the number of iterations for the 10 VQE results and plot the average in Fig.~\ref{fig:convergence_plots} (a). Similarly, in Fig.~\ref{fig:convergence_plots} (b), we plot the average number of iterations over the ten VQE results from~\ref{fig:diss} (a), for each bond length and each error mitigation technique. We find that as the coupling strength increases, the number of iterations required to converge decreases. We also find that using readout-error mitigation and ZNE together greatly reduces the required number of iterations, but this does not reduce the simulation time due to the overhead necessary to execute these mitigation techniques.

\section{Summary and Conclusion}

In this work, we show the viability of the PUCC ansatz for calculating the ground state energy by using the VQE algorithm on realistic noise models. Furthermore, we use the optimized parameters from the VQE to calculate the average photon number. The performance of this method is tested on an H$_2$ molecule interacting with a single photon in a single cavity mode. We use Qiskit's noisy simulators to obtain ground state energies across different regimes of bond lengths, cavity frequencies, and coupling strengths. Additionally, we employ various error mitigation techniques and show that the availability of reference states for the electron-photon problems offers unique advantages in achieving chemically accurate results. Our VQE results for the ground state energies are in excellent agreement with the FCI results, with many of the VQE results falling within chemical accuracy. Future work will include the calculation of excited state energies~\cite{ollitrault2020quantum,Pavosevic2021,pavosevic2023spinflip} using the quantum electrodynamic equation-of-motion (QED-qEOM) method which was developed alongside the PUCC ansatz~\cite{Pavosevic2021}. The methods presented in this work open up many additional research directions for developments in computational polaritonic quantum chemistry using quantum computing as well as in using quantum devices for strongly coupled fermion-boson systems.

\section*{Acknowledgements}
We acknowledge startup funding from the City College of New York. MH acknowledges funding from grant number EES-2112550 (NSF Phase II CREST Center IDEALS). The Flatiron Institute is a division of the Simons Foundation.

\bibliography{refs} 
\end{document}